\documentstyle[aps,preprint,tighten,epsf]{revtex}

\def\cm-3{\,{\rm cm}^{-3}}
\def\cos{{\rm cos}}
\def\df{{\rm d}}
\def\ev{\,{\rm eV}}
\def\gev{\,{\rm GeV}}
\def\halo{{\rm halo}}
\def\jet{{\rm jet}}
\def\kpc{\,{\rm kpc}}
\def\max{{\rm max}}

\def\mpc{\,{\rm Mpc}}
\def\therm{{\rm therm}}
\def\sec{\,{\rm sec}}
\def\yr{\,{\rm yr}}

\begin{document}
\draft 
\small
\preprint{\makebox{\begin{tabular}{r} 
 OUTP-97-60-P \\ hep-ph/9804285 \\ \end{tabular}}}
\title{Extremely high energy cosmic rays from relic particle decays}
\author{\large Michael Birkel \&\ 
         Subir Sarkar\thanks{s.sarkar@physics.ox.ac.uk} \\ \bigskip}
\address{Theoretical Physics, University of Oxford, \\ 
          1 Keble Road, Oxford OX1 3NP, UK \\ \bigskip}
\maketitle
\begin{abstract}
The expected proton and neutrino fluxes from decays of massive
metastable relic particles are calculated using the HERWIG QCD event
generator. The predicted proton spectrum can account for the observed
flux of extremely high energy cosmic rays beyond the
Greisen-Zatsepin-Kuzmin cutoff, for a decaying particle mass of ${\cal
O}(10^{12})$~GeV. The lifetime required is of ${\cal O}(10^{20})$~yr
if such particles constitute all of the dark matter (with a
proportionally shorter lifetime for a smaller contribution). Such
values are plausible if the metastable particles are hadron-like bound
states from the hidden sector of supersymmetry breaking which decay
through non-renormalizable interactions. The expected ratio of the
proton to neutrino flux is given as a diagonistic of the decaying
particle model for the forthcoming Pierre Auger Project.
\end{abstract}
\bigskip
\pacs{98.70.Sa, 14.80.-j}

\section{Introduction}

It has been known for some time that interactions on the 2.73 K
blackbody cosmic microwave background (CMB) will severely degrade the
energies of cosmic ray nucleons with energies beyond
$\sim5\times10^{19}\ev$ --- the Greisen-Zatsepin-Kuzmin (GZK) cutoff
\cite{gzk}. It was therefore very surprising when the Fly's Eye
atmospheric fluorescence detector reported the observation of an
extremely high energy cosmic ray (EHECR) event with an energy of
$(3.0\pm0.9)\times10^{20}\ev$ \cite{flyseye}. This was followed by the
detection of a $(1.7-2.6)\times10^{20}\ev$ event by the AGASA air
shower array \cite{agasa}. These discoveries substantiated earlier
claims from the Volcano Ranch \cite{vr}, Haverah Park \cite{hp} and
Yakutsk \cite{yak} air shower arrays that cosmic rays do exist beyond
the GZK cutoff. About a dozen such events are now known. Detailed
accounts of the data may be found in recent reviews \cite{reviews}.

\begin{figure}[t]
\epsfxsize\hsize\epsffile{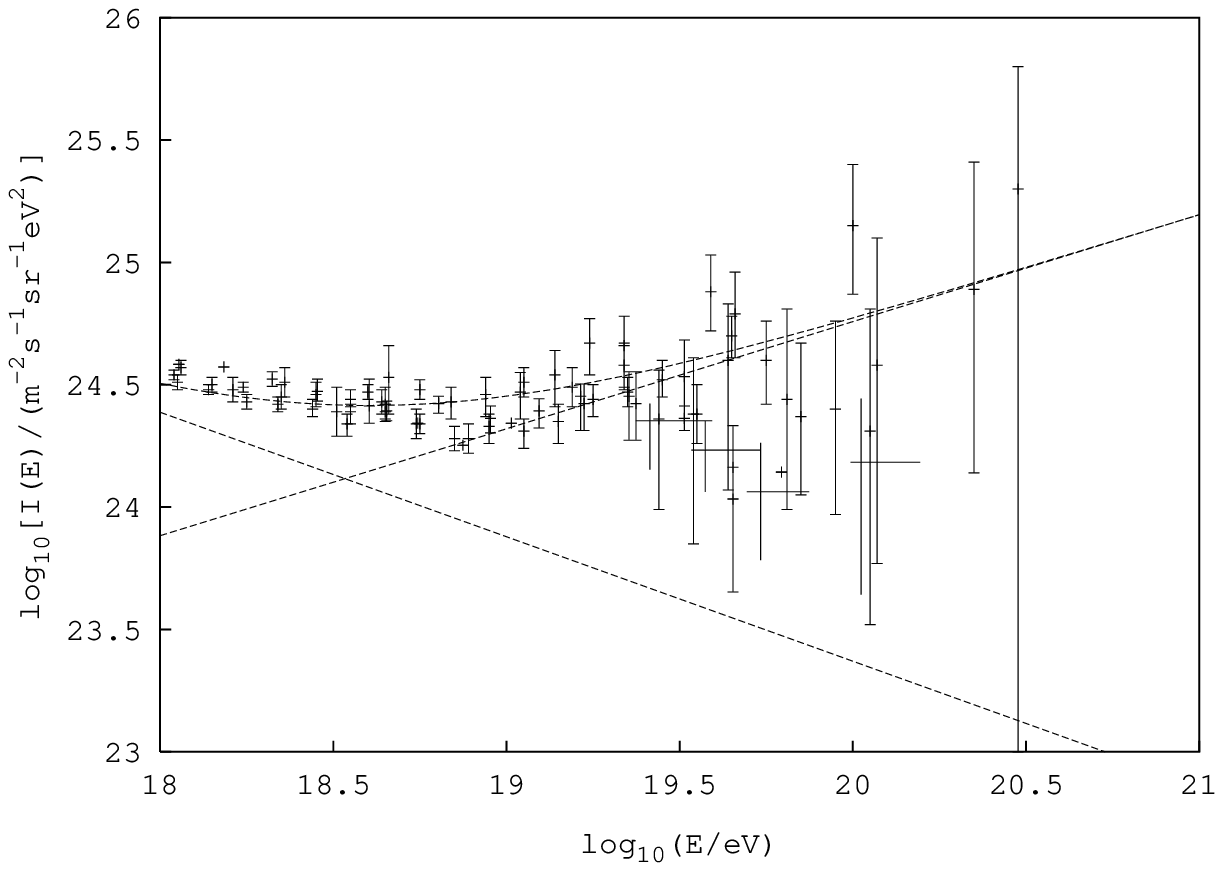}
\bigskip
\caption{The high energy cosmic ray spectrum beyond the `ankle'.
(Note that the differential flux has been multiplied by $E^3$.) The
data shown are from AGASA, stereo Fly's Eye, Haverah Park and Yakutsk
and are AGASA-normalized \protect\cite{spec}. The highest energy
monocular Fly's Eye event at $3\times10^{19}\ev$ is also shown. A fit
to the spectrum from the superposition of a steeply falling and a
flatter power law (dashed lines) is indicated \protect\cite{flyseye}.}
\label{fig1}
\end{figure}

In Figure~\ref{fig1} we show the EHECR spectrum for energies exceeding
$10^{18}\ev$ \cite{spec}; note that the fluxes have been multiplied by
$E^3$. It is believed that cosmic rays with energies up to
$\sim5\times10^{18}\ev$, the so-called `ankle', are predominantly of
galactic origin, possibly accelerated by the Fermi mechanism in
supernova remnants \cite{books}. Above this energy, the spectrum
flattens and the composition changes from being mostly heavy nuclei to
mostly protons. Such a correlated change in the spectrum and
composition was first established by the Fly's Eye experiment
\cite{flyseye} and Figure~\ref{fig1} shows their suggested
two-component fit to the data. The new component which dominates at
energies beyond $\sim5\times10^{18}\ev$ is isotropic and therefore
cannot possibly originate in the galactic disk
\cite{agasa2,lc95}. However it also extends well beyond the GZK cutoff
raising serious problems for hypothetical extragalactic
sources. Because of the rapid energy degradation at these energies
through photo-pion production on the CMB, such sources must exist
within $\sim500\mpc$, in fact within $\sim50\mpc$ for the highest
energy Fly's Eye event \cite{cronin}. For heavy nuclei, the energy
loss is less severe according to a revised calculation \cite{heavy} so
the range may extend upto $\sim100\mpc$.  General arguments
\cite{greisen,hillas} provide correlated constraints on the magnetic
field strength and spatial extent of the region necessary to
accelerate particles to such high energies and these requirements are
barely met by likely astrophysical sites such as active galactic
nuclei and the `hot spots' of radio galaxies \cite{cr}. Moreover there
are few such sources close to us and no definite correlations have
been found between their locations and the arrival directions of the
most energetic events \cite{source,agasa2}. It has been speculated
that gamma-ray bursts which too are isotropically distributed, may be
responsible for EHECRs \cite{grb}. However since these are at
cosmological distances, one would expect to see the GZK cutoff in the
cosmic ray spectrum contrary to observations (cf. ref.\cite{grbtest}).

Some of the above arguments may be evaded if the EHECR events are due
not to nucleons but neutral particles such as photons and neutrinos.
Although high energy photons also suffer energy losses in traversing
the CMB and the extragalactic radio background, there is no threshold
effect which would cause a cutoff near the GZK value
\cite{photon}. However the observed shower profile of the highest
energy Fly's Eye event \cite{flyseye} argues against the primary being
a photon since it would have interacted on the geomagnetic field and
started cascading well before entering the atmosphere
\cite{notphoton}. The observed events are also unlikely to be
initiated by neutrinos as they all have incident angles of less than
$40^\circ$ from the zenith and thus too small a path length in the
atmosphere for interactions \cite{gqrs96}. This argument may be evaded
if neutrinos become strongly interacting at high energies due to new
physics beyond the Standard Model \cite{neutrino,hongmo}, but such
proposals are found not to be phenomenologically viable \cite{bhg98}
(although this is disputed \cite{nu}). (Alternatively, the propagating
high energy neutrinos could annihilate on the relic cosmic neutrino
background, assumed to have a small mass of ${\cal O}(0.1)$~eV, to
make hadronic jets within the GZK zone \cite{weiler}.) Other exotic
possibilities have been suggested, e.g. monopoles \cite{wk96}, stable
supersymmetric hadrons \cite{farrar} and loops of superconducting
cosmic string (`vortons') \cite{bp97}. However these possibilities
have many phenomenological problems \cite{mn98,ber} and we do not
discuss them further.

Thus one is encouraged to seek `top-down' explanations for EHECRs in
which they originate from the decay of massive particles, rather than
being accelerated up from low energies. The most discussed models in
this connection are based on the annihilation or collapse of
topological defects such as cosmic strings or monopoles formed in the
early universe \cite{hill,td,bs95,bv97}. When topological defects are
destroyed their energy is released as massive gauge and Higgs bosons
which are expected to have masses of ${\cal O}(10^{16})\gev$ if such
defects have formed at a GUT-symmetry breaking phase transition. The
decays of such particles can generate cascades of high energy
nucleons, $\gamma$-rays and neutrinos. A more recent suggestion is
that EHECRs arise from the decays of metastable particles with masses
$m_X\sim10^{13}-10^{16}\gev$ which constitute a fraction of the dark
matter \cite{bkv97}. These authors suggest that such particles can be
produced during reheating following inflation or through the decay of
hybrid topological defects such as monopoles connected by strings, or
walls bounded by strings. The required metastability of the particle
is ensured by an unspecified discrete symmetry which is violated by
quantum gravity (wormhole) effects. Another suggestion is that the
long lifetime is due to non-perturbative instanton effects
\cite{kr97}. In ref.\cite{fkn97}, a candidate metastable particle is
identified in a $SU(15)$ GUT.

A generic feature of these `top-down' models is that the EHECR
spectrum resulting from the decay cascade is essentially determined by
particle physics considerations. Of course the subsequent propagation
effects have astrophysical uncertainties but since the decays must
occur relatively locally in order to evade the GZK cutoff
\cite{bkv97}, they are relatively unimportant. Thus although the
proposal is speculative, it is possible, in principle, to make
reliable calculations to confront with data. In this work we consider
another possible candidate for a relic metastable massive particle
\cite{ben98} whose decays can give rise to the observed highest energy
cosmic rays. First we discuss (\S~\ref{crypton}) why this candidate,
which arises from the hidden sector of supersymmetry breaking, is
perhaps physically better motivated than the other suggested
relics. We then undertake (\S~\ref{decay}) a detailed calculation of
the decay cascade using a Monte Carlo event generator to simulate
non-perturbative QCD effects. This allows us to obtain a more reliable
estimate of the cosmic ray spectrum than has been possible in earlier
work on both topological defect models \cite{td} and a decaying
particle model \cite{bkv97}. We confront our results with observations
and identify the mass and abundance/lifetime required to fit the
data. We conclude (\S~\ref{concl}) with a summary of experimental
tests of the decaying particle hypothesis.

\section{Massive, metastable dark matter from the hidden sector}\label{crypton}

Soon after the discovery of the anomaly-free heterotic superstring
theory in ten dimensions based on the gauge group $E_8\times\,E_8$, it
was pointed out \cite{ww85} that in the physical low energy theory
(where a grand unified $E_6$ or $O(10)$ group is broken by Wilson
lines), the minimum value of magnetic charge is not the Dirac quantum
$2\pi/e$ but an integral multiple thereof. Conversely, the minimum
electric charge is smaller than the electron charge $e$ by the same
ratio.
\footnote{The vacuum state of a physical theory in this scheme must be
$M^4\times\,K$ where $M^4$ is four-dimensional Minkowski space and $K$
is some compactified six-manifold. Such fractionally charged states
exist because $K$ is not simply connected --- these are states in
which a closed string wraps around a non-contractible loop in $K$.}
This was found to be a generic feature of all superstring models based
on a level-one Ka\u{c}-Moody algebra \cite{level1}. In view of the severe
experimental upper bounds on the relic abundance of fractional charges
\cite{pdg}, this posed a potential embarrassment for superstring
phenomenology \cite{frac}.

A simple solution to the problem of fractional charges (with an
obvious historical analogue in quarks and QCD) is to confine them and
it was shown that this can be done in the hidden sector of
supersymmetry breaking in the framework of the $SU(5)\otimes\,U(1)$
unification model \cite{aehn89}. In this model, all fractionally
charged states have charges $|Q_{\rm em}|=\frac{1}{2}$ and are placed
in {\bf 4} or {\bf 6} representations of a hidden $SU(4)$ gauge group
which becomes strong at a scale $\Lambda_4\sim10^{12}\gev$ and in {\bf
10} representations of a hidden $SO(10)$ group which becomes strong at
a scale $\Lambda_{10}\sim10^{15}\gev$. This results in integer-charged
particles --- `cryptons' --- which may be 2-constituent mesons,
3-constituent baryons or 4-constituent `tetrons' \cite{eln90}. Some of
these mesons could be light (in analogy to the pion of QCD) but most
of the states should be heavy with masses of order the confinement
scale $\Lambda$. (Other possibilities for stable superstring relics
have been discussed in ref.\cite{exotic}.)

The constituent fields have very few renormalizable ($N=3$)
superpotential interactions, so most of these states can only decay
via higher-order ($N\geq4$) superpotential terms. Generically, crypton
lifetimes are expected to be \cite{eglns92}
\begin{equation}
 \tau_X \simeq {1 \over m_X} \left(\frac{M}{m_X}\right)^{2(N-3)},
  \quad m_X \sim \Lambda, 
\label{taucrypton}
\end{equation}
where, $M\equiv\,M_{\rm P}/\sqrt{8\pi}\simeq2.4\times10^{18}$~GeV is the
normalized Planck scale, giving
\begin{equation}
\tau_4 \sim 10^{(12N_4 - 80)} \yr, \quad 
 \tau_{10} \sim 10^{(6N_{10} - 65)} \yr,
\label{tau10tau4}
\end{equation}
for $SU(4)$ and $SO(10)$ bound states respectively. Thus $\tau_4
\gtrsim\,1\,\sec\,(10^{16}\yr)$ for $N_4\geq\,6\,(8)$ and
$\tau_{10}\gtrsim\,1\,\sec\,(10^{16}\yr)$ for $N_{10}\geq10\,(14)$.
Detailed studies of the possible effects of decays of relic cryptons
on primordial nucleosynthesis and the CMB
spectrum \cite{eglns92}, as well as on the diffuse $\gamma$-ray
background \cite{eglns92,kr297} have established that such particles,
if they survive as relics of the Big Bang, must either decay well
before nucleosynthesis or have lifetimes longer than the age of the
universe ($t_0\sim10^{10}\yr$). In the latter case, if such particles
make an interesting contribution to the dark matter, their lifetime is
further required to exceed $\sim10^{16}\yr$ in order to respect
experimental bounds on the flux of high energy neutrinos expected from
their decays \cite{eglns92,ggs93}. It is seen from
eq.(\ref{tau10tau4}) that these constraints favour $SU(4)$ mesons over
their $SO(10)$ counterparts as possible constituents of the dark
matter. It is then natural to contemplate the possibility that such
cryptons with a mass of $m_4\sim10^{12}\gev$ and a lifetime
$\tau_4\gtrsim10^{16}\yr$ are also responsible for the observed
highest energy cosmic rays.

Recently the above discussion has been extended to other massive
metastable particle candidates in superstring/M-theory
\cite{ben98}. These authors discuss constructions with higher-level
Ka\u{c}-Moody algebras (necessary to accommodate adjoint Higgs
representations in (unified) models other than $SU(5)\otimes\,U(1)$)
and note that similar metastable bound states occur in such
models. They go on to consider other candidate particles in M-theory
such as Kaluza-Klein states associated with extra dimensions but find
that these are not as attractive, being either too heavy or too
unstable. They suggest that although the $SU(5)\times\,U(1)$ model
\cite{aehn89} discussed above was constructed in the weak coupling
limit, it may be elevated to an M-theory model in the strong coupling
limit. The $SU(4)$ tetrons are then still the most likely candidates
for massive metastable dark matter with the modification that the
Planck scale $M$ in Eq.(\ref{taucrypton}) may be replaced by a
somewhat smaller scale.

The main reason why this possibility was not seriously entertained
earlier concerns the expected relic abundance of such massive
particles. If cryptons were maintained in chemical equilibrium in the
early universe through self-annihilations, their present energy
density is given by the usual `freeze-out' calculation as inversely
proportional to the (velocity-averaged) annihilation cross-section
\cite{relic}. Estimating this to be $\langle\sigma_{\rm
ann}v\rangle\sim\,m_X^{-2}$ we see that equilibrium would have been
established if the annihilation rate exceeded the Hubble expansion
rate ($H\sim\,T^2/M$), i.e. at temperatures
\begin{equation}
\label{decoup}
  T > T_{\rm dec} \sim \frac{m_X}{\ln(M/m_X)} .
\end{equation}
The relic abundance is then simply estimated as the equilibrium value
at decoupling:
\begin{equation}
\label{crypdens}
  \Omega_X \sim 10^{14} \left(\frac{m_X}{10^{12}\gev}\right)^2 .
\end{equation}
This is the basis for the conclusion that no stable relic particle may
have a mass in excess of $\sim10^5\gev$ without `overclosing' the
universe, i.e. contributing $\Omega_X>1$ \cite{relic,exotic}. This
does not necessarily apply to cryptons since a period of inflation
should have diluted their abundance to essentially zero, along with
monopoles and other such supermassive relics. If the reheating
temperature following inflation is restricted to be $T_{\rm
R}\lesssim10^{9}-10^{10}\gev$ in order not to produce too many
gravitinos \cite{ekn84,sark96}, cryptons would not have been generated
afterwards.

However it has been recently recognized that in supersymmetric
cosmology, there is likely to be a late stage of `thermal inflation'
\cite{thermal} due to symmetry breaking along flat directions at
intermediate scales \cite{crisis,interm}.
\footnote{This was initially considered to be an `entropy crisis'
\cite{crisis} since it would dilute any baryon asymmetry generated at
the GUT scale. However there are now several plausible mechanisms for
low temperature baryogenesis \cite{electrobaryo,susybaryo} which may
operate after thermal inflation.}
This would adequately dilute the abundance of thermally
generated gravitinos following inflation so the bound quoted above on
the reheating temperature is no longer valid and the value of $T_{\rm
R}$ may be much higher.
\footnote{The vacuum energy $V(\phi)$ of the scalar `inflaton' field
is constrained to be $V^{1/4}/\epsilon^{1/4}\simeq2.7\times10^{-2}M$
by the anisotropy in the CMB observed by COBE, where the slope
parameter $\epsilon\equiv\,(M^2/2)(V'/V)^2$ is required to be $\ll1$
to permit inflation to occur \cite{infl}. (The number of e-folds of
expansion until the end of inflation is just $N=\int^{\phi_{\rm
end}}_{\phi}\df\phi/M\sqrt{2\epsilon}$ and this should exceed
$\sim50-60$ in order to solve the flatness and homogeneity problems of
the standard cosmology.) The reheat temperature $T_{\rm R}$ can, in
principle, have been as high as $\sim\,V^{1/4}$ although it is usually
considerably smaller since the inflaton field is very weakly coupled
in most inflationary models.}
In that case cryptons even as massive as $10^{12}\gev$ may well have
been brought back into thermal equilibrium during reheating after
inflation and survived with the huge relic abundance
(\ref{crypdens}). However thermal inflation would also have diluted
this to an acceptable level as was noted in ref.\cite{ars97}; to
obtain $\Omega_X\sim1$, the number of e-folds of thermal inflation
required is just
\begin{equation}
    N_\therm = \frac{1}{3} \ln (10^{14}) \simeq 11 .
\end{equation}
This fits in well with the expectation that
$N_\therm\sim\case{1}{2}\ln(\Sigma/m_W)\sim10-15$ for the intermediate
scale $\Sigma$ in the range $(10^{-7}-10^{-2})M$ \cite{thermal}. Of
course given the uncertainty in the value of $N_\therm$ (and indeed
the possibility that there may be more than one such epoch),
$\Omega_X$ could well have been reduced to a negligibly small value.

Another possibility is that massive particles such as cryptons were
never in thermal equilibrium but were created with a cosmologically
interesting abundance due to the varying gravitational field during
(primordial) inflation \cite{ckr98,kt98}. A cosmologically interesting
relic abundance then arises for $m_X\sim(0.04-2)H$ where
$H\sim10^{11}-10^{14}\gev$ is the likely Hubble parameter during
inflation \cite{ckr98}. This is certainly very encouraging but it
should be remembered that a later stage of thermal inflation would
dilute such an abundance to a negligible level, as discussed above.

It is clear that the relic abundance of massive particles such as
cryptons will necessarily be very uncertain given our ignorance of the
thermal history of the universe prior to nucleosynthesis. However as
the above discussion illustrates, there are two complementary ways in
which a cosmologically interesting abundance may result so we may
reasonably consider such particles as candidates for the dark
matter. We now move on to discuss whether relic cryptons can indeed be
the source of the EHECR by determining the expected spectrum of high
energy particles from their decays.

\section{Cosmic rays from massive particle decay}\label{decay}

To calculate the expected flux of cosmic rays from decays of massive
particles such as cryptons, we must consider the contribution from
both decaying particles in the halo of our Galaxy as well as those
elsewhere in the universe. Since such massive particles would behave
as cold dark matter and cluster efficiently in all gravitational
potential wells, their abundance in our galactic halo would be
enhanced above their cosmological abundance by a factor
\begin{equation}
   f_\cos \equiv n_X^\halo/n_X^\cos .
\end{equation}
Note that $\Omega_X=m_X\,n_X^\cos/\rho_{\rm crit}$ where $\rho_{\rm
crit}\simeq1.054\times10^{-4}h^2\gev\,\cm-3$ is the critical density
in terms of the present Hubble parameter
$h\equiv\,H_0/100$\,km\,sec$^{-1}$\,Mpc$^{-1}$. If for simplicity we
assume a spherical halo of uniform density,
\begin{equation}
 R_\halo \sim 100 \kpc, \qquad \rho^\halo \sim 0.3 \,\gev\,\cm-3,
\end{equation}
then $f_\cos\sim3\times10^4h^{-2}$ and the number density of cryptons
in the halo is
\begin{equation}
\label{cryptdens}
  n_X^\halo \sim 3 \times 10^{-13} \cm-3\ 
   \left(\frac{f_\cos \Omega_X h^2}{3 \times 10^4}\right)\  
   \left(\frac{m_X}{10^{12}\gev}\right)^{-1} .
\end{equation}
The actual density of dark matter in the halo must of course fall off
as $r^{-2}$ to account for the flat rotation curve of the disk but we
do not consider it necessary at this stage to investigate realistic
mass models. Thus the universal density of cryptons is smaller than
the halo density by about the same numerical factor by which the
distance to the horizon ($\sim3000h^{-1}\mpc$) exceeds the halo radius, so
the extragalactic contribution to the EHECR flux from decaying
cryptons cannot exceed the halo contribution. In particular, the GZK
cutoff scale for protons \cite{cronin} or heavy nuclei \cite{heavy} are
all much smaller than the horizon distance, so only the halo
contribution need be considered, as was emphasized in
ref.\cite{bkv97}. Only for neutrinos would the extragalactic component
be comparable in magnitude \cite{ggs93}. Henceforth we restrict
ourselves to considering crypton decays in the halo alone.

Now the injection spectrum from particle decay is, to a good
approximation,
\begin{equation}
\label{injspec}
 \Phi_i = \left|\frac{\df n_X}{\df t}\right| \frac{\df N_i}{\df E} =
	\frac{n_X^\halo}{\tau_X} \frac{2}{m_X} \frac{\df N_i}{\df x}
\end{equation}
for lifetimes longer than the age of the universe
($\tau_X\gg\,t_0$). Here
\begin{equation}
\label{x}
   x = \frac{E}{E_\jet} = \frac{2 E}{m_X} 
\end{equation}
is a measure of particle energy (assuming 2-body decays) and the
fragmentation function $\df\,N_i/\df\,x$ is the average number of
particles $i$ released per decay, per unit interval of $x$, at the
value $x$. The flux at Earth is then
\begin{equation}
\label{flux}
   j_i (E) = \frac{1}{4\pi} R_\halo \Phi_i (E) .
\end{equation}

The final state particles which interest us most are `protons' and
neutrinos/antineutrinos where the former includes other nucleons,
e.g. antiprotons and neutrons, since they all interact similarly in
the Earth's atmosphere. To compare with observations we multiply the
fluxes by $E^3$ and define
\begin{eqnarray}
\label{fluxI}
   I_p(E) & \equiv j_p (E) E^3 
   & = \frac{1}{4\pi} \frac{n_X^\halo}{\tau_X} 
    R_\halo \frac{2}{m_X} \frac{\df N_p}{\df x} E^3  , \\ \nonumber
   I_\nu (E) & \equiv j_\nu (E) E^3
   & = \frac{1}{4\pi} \frac{n_X^\halo}{\tau_X} 
    R_\halo \frac{2}{m_X} \frac{\df N_\nu}{\df x} E^3 .
\end{eqnarray}
For photons and electrons/positrons, propagation energy losses are
substantial even within the halo and we do not attempt to determine
these. However their injection spectra from particle decay are given
by the same computation as for protons and neutrinos, to which we now
turn.

\subsection{Computing the fragmentation functions}

Heavy particles, whether GUT-scale bosons (in topological defect
models), cryptons or other hypothetical massive particles, will decay
into quarks and leptons. The quarks will hadronize producing jets of
mostly pions with a small admixture of nucleons and antinucleons. The
neutral pions will decay to give photons while charged pion decays
will yield neutrinos and antineutrinos in addition to leptons. Thus
the final spectrum of the decay produced particles will be essentially
determined by the `fragmentation' of quarks/gluons into hadrons. This
is a non-perturbative QCD process and it has not been possible to
calculate it by analytic means. Usually phenomenologically motivated
approximations are used to model experimental data on inclusive jet
multiplicities and scaling violations \cite{frag}.

So far, authors of proposals involving heavy particle decay, e.g. in
the context of topological defect models \cite{td}, have employed a
hadronic fragmentation function suggested by Hill \cite{hill}
\begin{equation}
\label{hillfrag}
 \frac{\df N_h^{\rm (Hill)}}{\df x} = 0.08 
   \frac{\exp[2.6\sqrt{\ln(1/x)}](1-x)^2}{x\sqrt{\ln(1/x)}} .
\end{equation} 
It is further {\em assumed} that 3\% of the hadronic jets from massive
relic particle decays turn into nucleons, while the other 97\% are
pions which decay into photons and neutrinos. This was based on the
leading logarithm approximation of QCD \cite{leadlog} applied to
experimental data from PETRA on jet production in $e^+e^-$ collisions
at tens of GeV. The estimated jet multiplicity from gluon
fragmentation was convoluted with the gluon distribution to determine
the total hadron yield; to estimate the spectrum, it was assumed that
the first moment of the distribution is normalized to unity and the
large $x$ behaviour was guessed to be $(1-x)^2$ \cite{hill}. As we
shall see, the Hill fragmentation function (\ref{hillfrag})
significantly {\em overestimates} the yield of high $x$ final states
from the decay of very massive particles and, moreover, photons and
neutrinos are actually produced with a spectrum quite different from
that of nucleons. Thus the decay spectra derived using
eq.(\ref{hillfrag}) for topological defect models \cite{td} are
inaccurate.

Subsequently, another form called the Modified Leading Logarithm
Approximation (MLLA) which gives a better description of data at low
$x$ has been proposed \cite{frag}; a gaussian approximation to this is
\begin{equation}
\label{mlla}
 \frac{\df N_h^{\rm (MLLA)}}{\df x} = \frac{K_N}{x} \exp
     \left[\frac{\ln^2(x/x_\max)}{2\sigma^2}\right] ,
\end{equation}
where $K_N$ is a constant and 
\begin{equation}
 2\sigma^2 = \frac{\pi}{21} \sqrt{\frac{2\pi}{3\alpha_{\rm S}^3(s)}}
             \simeq 0.09 \left[ \ln \left(\frac{m_X^2}{\Lambda^2} 
             \right)\right]^{3/2} ,
\end{equation}
with $x_\max=\sqrt{\Lambda/m_X}$ and $\Lambda=0.234\gev$. This
fragmentation function is employed by the authors of ref.\cite{bkv97}
to compute the spectrum from relic particle decays; they determine
$K_N$ by requiring that the integral of $x\,{\rm d}\,N_h^{\rm
(MLLA)}/{\rm d}\,x$ over the range $x\in[0,1]$ be equal to the
fraction of the energy transferred to hadrons. However this procedure
is not exact as the form (\ref{mlla}) is inapplicable for large $x$
and therefore cannot be normalized in this manner. Thus the shape of
the cosmic ray spectrum computed \cite{bkv97} by this method for
decaying particles is only reliable for small $x$ and its
normalization uncertain.

Given the importance of determining the energy spectrum accurately, we
decided to improve on these approximate formulations by using the
standard tool employed by experimental high energy physicists to study
QCD fragmentation, viz. a Monte Carlo event generator. Here the
non-perturbative hadronization process is simulated on a computer by a
well tested phenomenological model \cite{evntgen}. Although this
requires extensive numerical calculations, it is the only means by
which successful contact can be made between theory and experimental
data. We chose the programme HERWIG \cite{herwig} (Hadron Emission
Reactions With Interfering Gluons) which incorporates the cluster
model for hadronization and is based on a shower algorithm of the `jet
calculus' type \cite{evntgen}. To check our results we also ran the
JETSET programme \cite{jetset} and found good agreement over the
energy range where comparison was possible. However for the very high
energies studied in this work, HERWIG proved to be more suitable for
reasons of computing time \cite{privcom}. Even so the calculations
described here took several months on a Digital Alpha workstation.

For definiteness, we assume the heavy particles to decay into a
quark-antiquark pair with unit branching ratio. The quark and
antiquark, each carrying away energy $m_X/2$, form jets which lead to
the generation of many particles through cascading, hadronization and
decays of some of the generated particles. This can be simulated by
HERWIG via the annihilation process $e^+ e^-\rightarrow\,q\bar{q}$
with center-of-mass energy $\sqrt{s}=m_X$, where $q$ stands for all
six kinematically allowed quark flavours. The event generator outputs
kinematical details of all final state particles, e.g. protons,
photons and leptons (electrons, positrons and neutrinos). We divided
the $x$-range into 100 bins of width $\Delta\,x=0.01$. After each
event simulation the number of protons, neutrinos, photons as well as
electrons and positrons per energy bin was counted. We ran 10000
events for each of the masses $m_X = 10^{3}, 10^{5}, 10^{7}, 10^{9}$
and $10^{11}\gev$. After all events had been run, the particle numbers
in the bins are divided by the bin width and the number of events, in
order to obtain the fragmentation functions $\df\,N_i/\df\,x$. Apart
from altering some relevant parameters in the computer code to allow
it to run at the high energies studied here, we also switched off
initial state radiation since it is not relevant for the present
study. Unfortunately, it was not feasible to study the high $x$
behaviour of the fragmentation functions for decaying particle masses
higher than $10^{11}\gev$ because of numerical convergence problems in
the computer code. (Already for masses exceeding $10^{9}\gev$
quadruple precision had to be used.) Hence we have had to extrapolate
the fragmentation functions to high $x$ for very heavy masses as
described later.

\begin{figure}[t]
\bigskip
\epsfxsize\hsize\epsffile{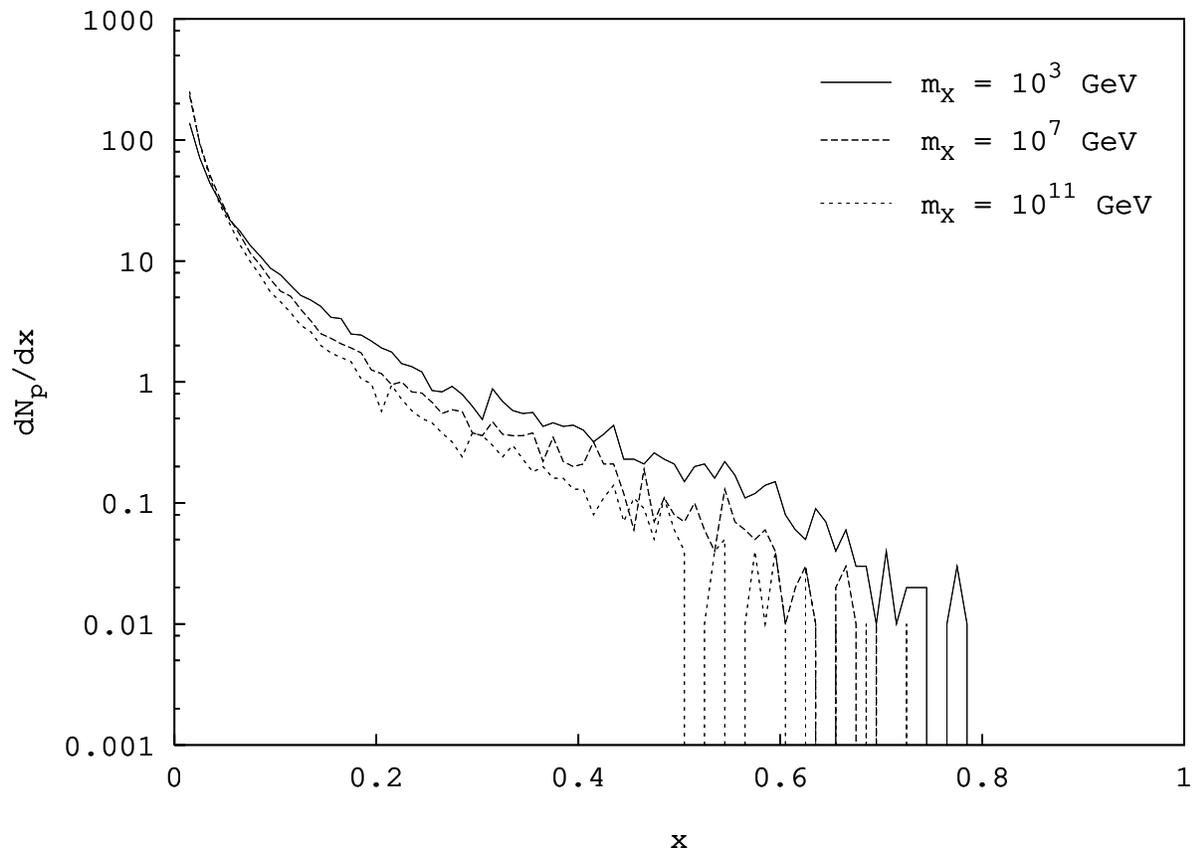}
\caption{Fragmentation function of protons, for decaying particle
masses $m_X=10^{3}$, $10^{7}$, and $10^{11}\gev$, computed using the
event generator HERWIG.}
\label{fig2}
\end{figure}

First we show the proton fragmentation function obtained from the
HERWIG runs in Figure~\ref{fig2} to illustrate that it depends on the
decaying particle mass, contrary to the approximation (\ref{hillfrag})
employed in previous work on topological defects \cite{td}. Rather
than being constant, it decreases with increasing $m_X$ for
$x\gtrsim0.1$, while at very low $x$ it increases with increasing
$m_X$. The large fluctuations at $x\gtrsim0.5$ are due to the fact
that relatively few particles are produced with such high energies
despite the 10000 events per simulation. We note also that the shape
differs significantly at high $x$ from the approximation used in
ref.\cite{bkv97}.

\begin{figure}[t]
\epsfxsize\hsize\epsffile{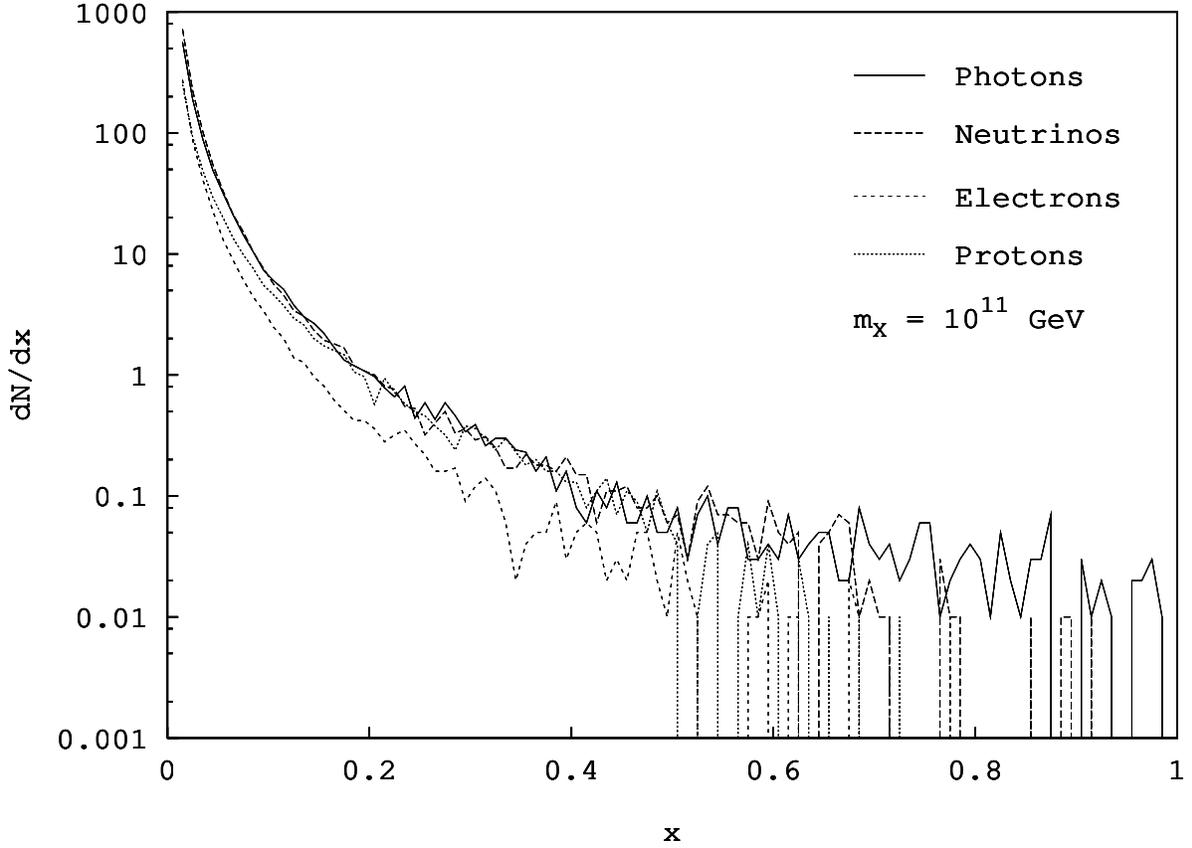}
\bigskip
\caption{Fragmentation functions of photons, neutrinos, and electrons
compared to that of protons for a decaying particle mass of
$m_X=10^{11}\gev$.}
\label{fig3}
\end{figure}

In Figure~\ref{fig3} the fragmentation functions for protons, photons,
neutrinos and electrons are compared for $m_X=10^{11}\gev$. It is seen
that at very low $x$ there are more photons and neutrinos generated by
the particle decay than electrons and protons. In the regime
$0.2\lesssim\,x\lesssim0.4$, photons, neutrinos and protons are
generated with roughly equal abundances. However, for $x\gtrsim0.5$,
photons and neutrinos again outnumber protons, in particular protons
cut off at $x\approx0.75$ whereas neutrinos and photons are generated
in the cascades with energies up to $x\approx0.95$. These differences
will lead to different shapes of the expected fluxes $I_i(E)$ as can
be seen from eq.(\ref{fluxI}).

\begin{figure}[t]
\epsfxsize\hsize\epsffile{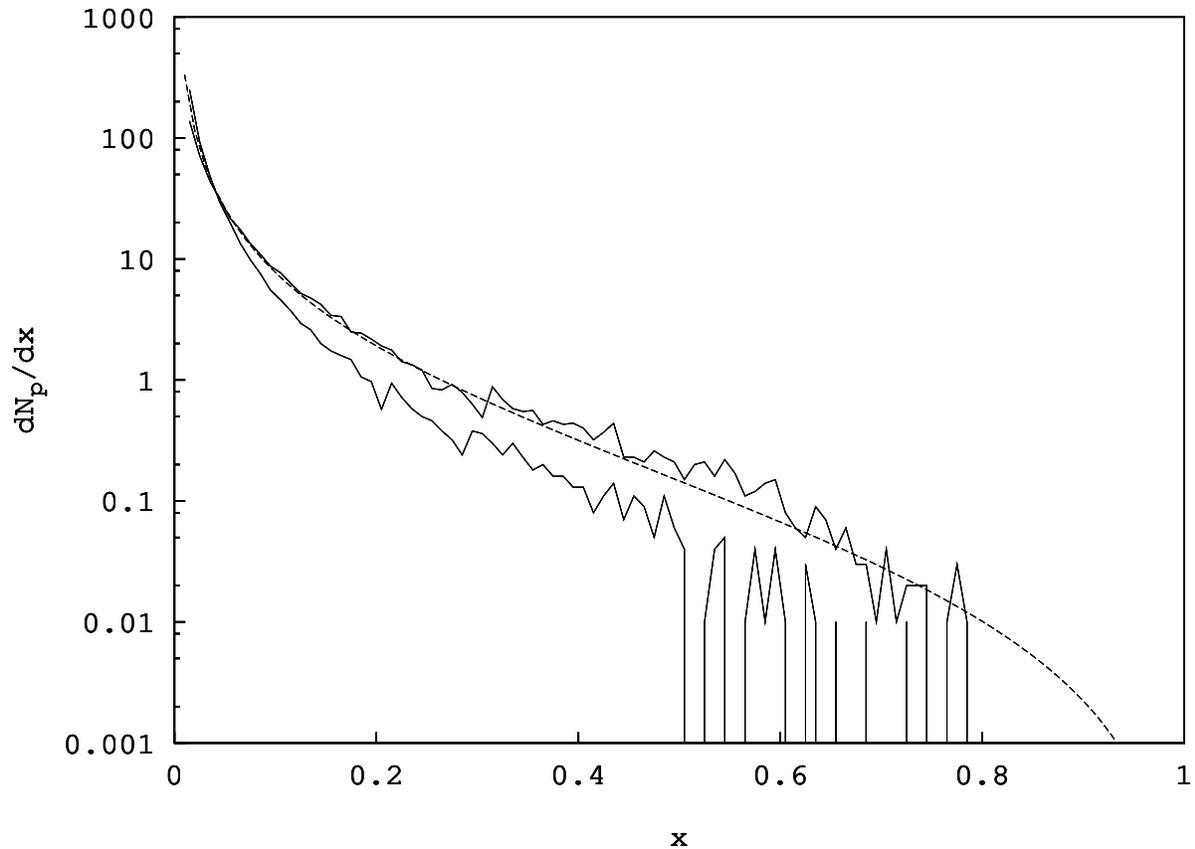}
\bigskip
\caption{Comparison of the computed proton fragmentation function with
the leading-log approximation of eq.(\protect\ref{hillfrag}),
normalized to the HERWIG computation at $x=0.042$. The upper solid
line refers to a decaying particle mass of $m_X=10^3\gev$ and the
lower one to $m_X=10^{11}\gev$.}
\label{fig4}
\end{figure}

We now compare our proton fragmentation function with the commonly
used Hill approximation \cite{hill} in Figure~\ref{fig4}. Although his
form provides a good fit for a low decaying particle mass,
viz. $m_X=10^3\gev$, it no longer does so for a high mass,
viz. $m_X=10^{11}\gev$. This is understandable given that the
numerical co-efficients in eq.(\ref{hillfrag}) were chosen to match
relatively low energy collider data. However the functional form
itself is well motivated and using our HERWIG runs we can determine
new numerical co-efficients appropriate to heavier mass particles.
Another advantage of the present approach is that the spectrum of
neutrinos and photons is determined separately from that of the
protons and not simply assumed to be proportional as in previous work
\cite{td,bkv97}.

\begin{figure}[t]
\epsfxsize\hsize\epsffile{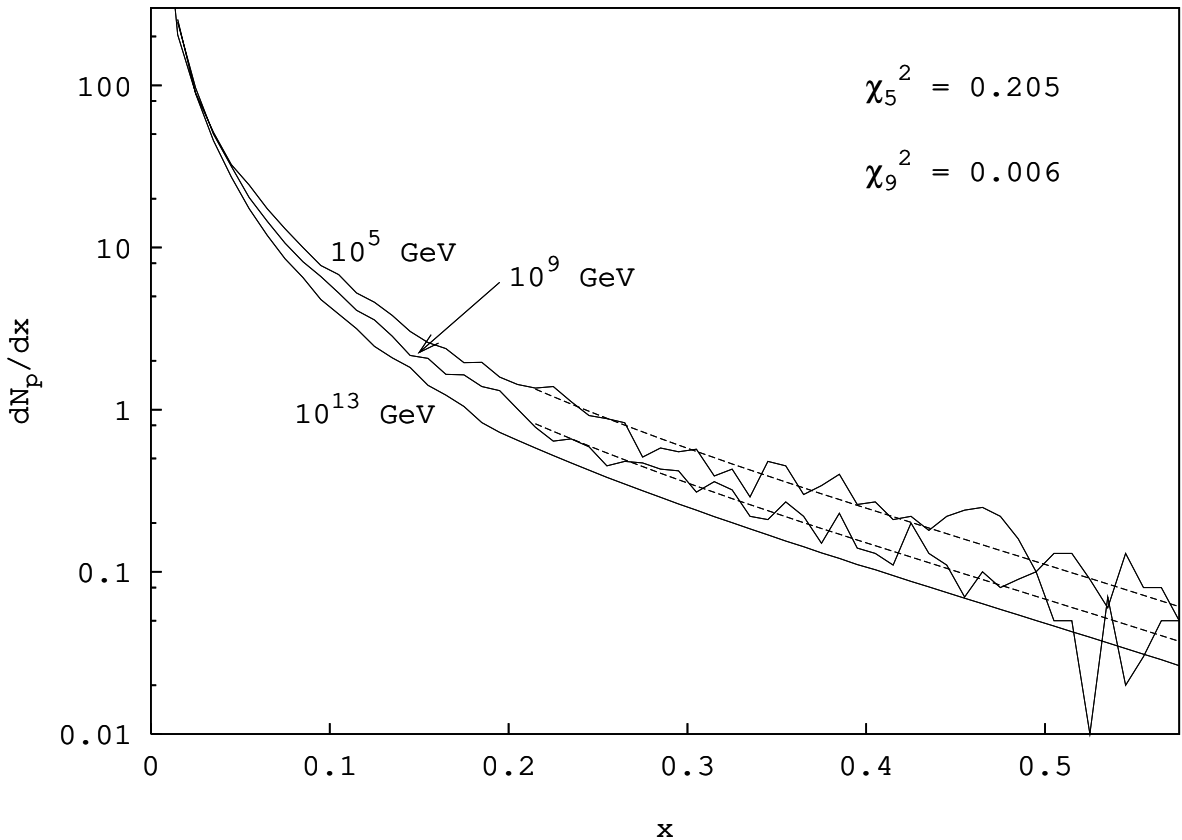}
\bigskip
\caption{Extrapolation of the computed proton fragmentation function
to higher decaying particle masses. The upper two solid lines are
HERWIG results for decaying particles of mass $m_X=10^5$ and
$10^9\gev$ while the dashed lines are the best fitting functions
according to eq.(\protect\ref{fit1}). (The corresponding
$\chi^2$-values are also indicated.) The lower solid line is the
fragmentation function for $m_X=10^{13}\gev$ obtained from
extrapolating the fitting parameters.}
\label{fig5}
\end{figure}

\begin{figure}[t]
\epsfxsize\hsize\epsffile{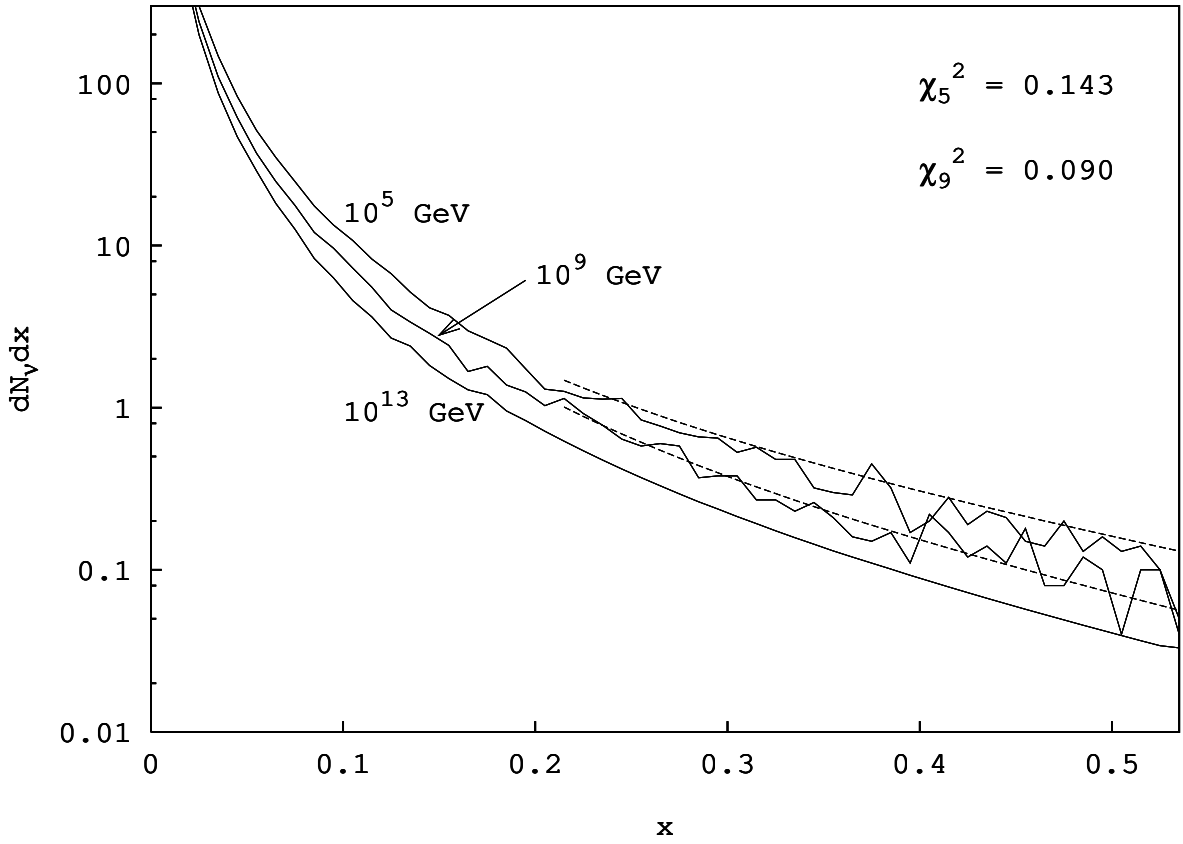}
\bigskip
\caption{Same as Figure~\protect\ref{fig5} for the case of neutrinos
with the fitting function now given by eq.(\protect\ref{fit2}).}
\label{fig6}
\end{figure}

To study the highest energy cosmic ray events we need to consider
particle masses beyond $10^{11}\gev$ but this is difficult to do
directly with HERWIG for technical reasons as mentioned earlier. We
therefore resort to an extrapolation procedure as follows. For the
range $x\in[0,0.2]$ the fragmentation functions are smooth and evolve
monotonically with $m_X$ so the fragmentation functions for a
$10^{13}\gev$ particle is obtained from simple linear extrapolation of
the lower energy fragmentation functions in each individual energy
bin. For $x\in[0.2,0.6]$ we first fit the calculated fragmentation
functions to the form
\begin{equation}
\label{fit1}
 \frac{\df\,N_{\rm fit1}}{\df\,x} = 
  c_1 \frac{\exp[c_2\sqrt{\ln(1/x)}](1-x)^2}{x\sqrt{\ln(1/x)}} ,
\end{equation} 
for protons, and the form 
\begin{equation}
\label{fit2}
  \frac{\df\,N_{\rm fit2}}{\df\,x} = 
  d_1 \frac{\exp[d_2\ln(1/x)](1-x)^2}{x\ln(1/x)} , 
\end{equation} 
which proves more suitable for photons, neutrinos and electrons. The
numerical co-efficients $c_1, c_2$ and $d_1, d_2$ are determined for
particle masses less than $10^{11}\gev$ by minimizing $\chi^2$ in the
fit to the actual HERWIG runs. In Figures \ref{fig5} and \ref{fig6} we
show these fits for $x\lesssim0.6$ to the proton and neutrino
fragmentation functions corresponding to masses of $10^5$ and
$10^9\gev$. Then we determine the appropriate co-efficients for
heavier masses by extrapolation. An example, for $m_X=10^{13}\gev$, is
shown in the figures. For $x\gtrsim0.6$, statistical fluctuations
become too severe so we extrapolate the fitting functions between the
value at $x=0.55$ and a cutoff which is taken to be $x=0.75$ for
protons and $x=0.95$ for neutrinos, based on the observed behaviour
for masses upto $10^{11}\gev$ shown in Figures~\ref{fig2} and
\ref{fig3}.

Finally, we mention the continuation of the proton fragmentation
function for very low $x$, viz. $x\lesssim0.015$, which is relevant at
high masses e.g. $m_X=10^{13}\gev$. Since it proved impractical to
have additional binning intervals at very small $x$, we employ the
fragmentation function (\ref{mlla}) in this regime, normalized to our
computations at $x=0.015$.

\subsection{Comparison with observations}

With the fragmentation functions obtained above, we can now calculate
the expected fluxes of protons and neutrinos from decays of particles
such as cryptons in the halo. We normalize the calculated proton flux
(\ref{fluxI}) to the observed cosmic ray flux at $10^{19}\ev$
\cite{flyseye}:
\begin{equation}
\label{FEnorm}
 \log_{10}[I_p (10^{19}\ev)/{\rm m}^{-2}\sec^{-1}{\rm sr}^{-1}\ev^2]
 = 24.32 .
\end{equation}
Note that the corresponding neutrino flux $I_\nu(E)$ is then a {\em
prediction} as the fragmentation function for neutrinos is computed
independently.

\begin{figure}[t]
\epsfxsize\hsize\epsffile{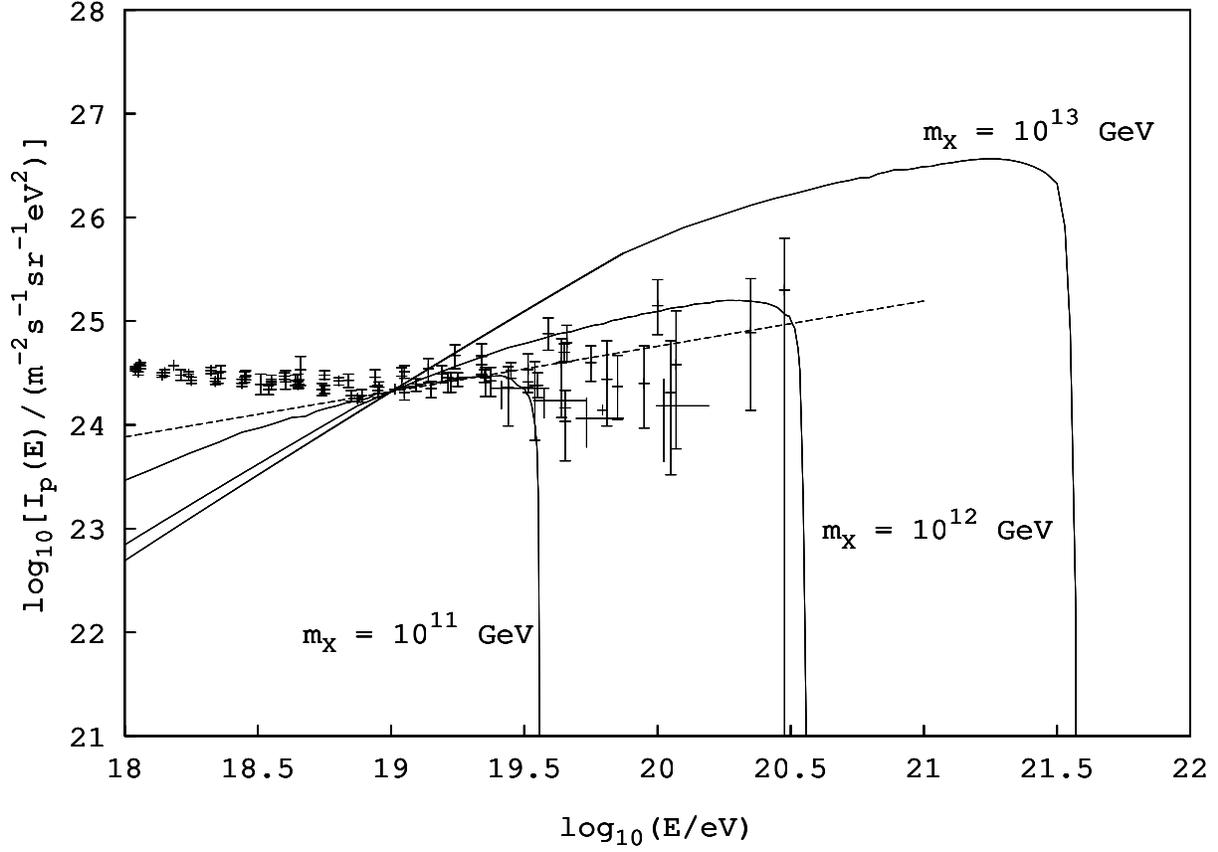}
\bigskip
\caption{Expected proton flux for decaying particle masses
$m_X=10^{11}$, $10^{12}$ and $10^{13}\gev$ compared with
observations. The theoretical spectra are normalized at $10^{19}\ev$
to the flat component (dashed line) suggested by the Fly's Eye data.}
\label{fig7}
\end{figure}

The expected proton fluxes are shown in Figure~\ref{fig7}. We see that
a crypton with $m_X=10^{11}\gev$ fits the flat power law well but
cannot explain the events beyond $4\times10^{19}\ev$. Although this is
easily achieved for $m_X=10^{13}\gev$, the decays of such a massive
particle would overproduce protons for
$E\gtrsim3\times10^{19}\ev$. Thus a crypton with mass
$m_X\sim10^{12}\gev$ provides the best compromise although it too
predicts a spectrum somewhat flatter than the one indicated
observationally. (The reader is reminded that all differential fluxes
have been multiplied by $E^3$ in eq.(\ref{fluxI}).)

\begin{figure}[t]
\epsfxsize\hsize\epsffile{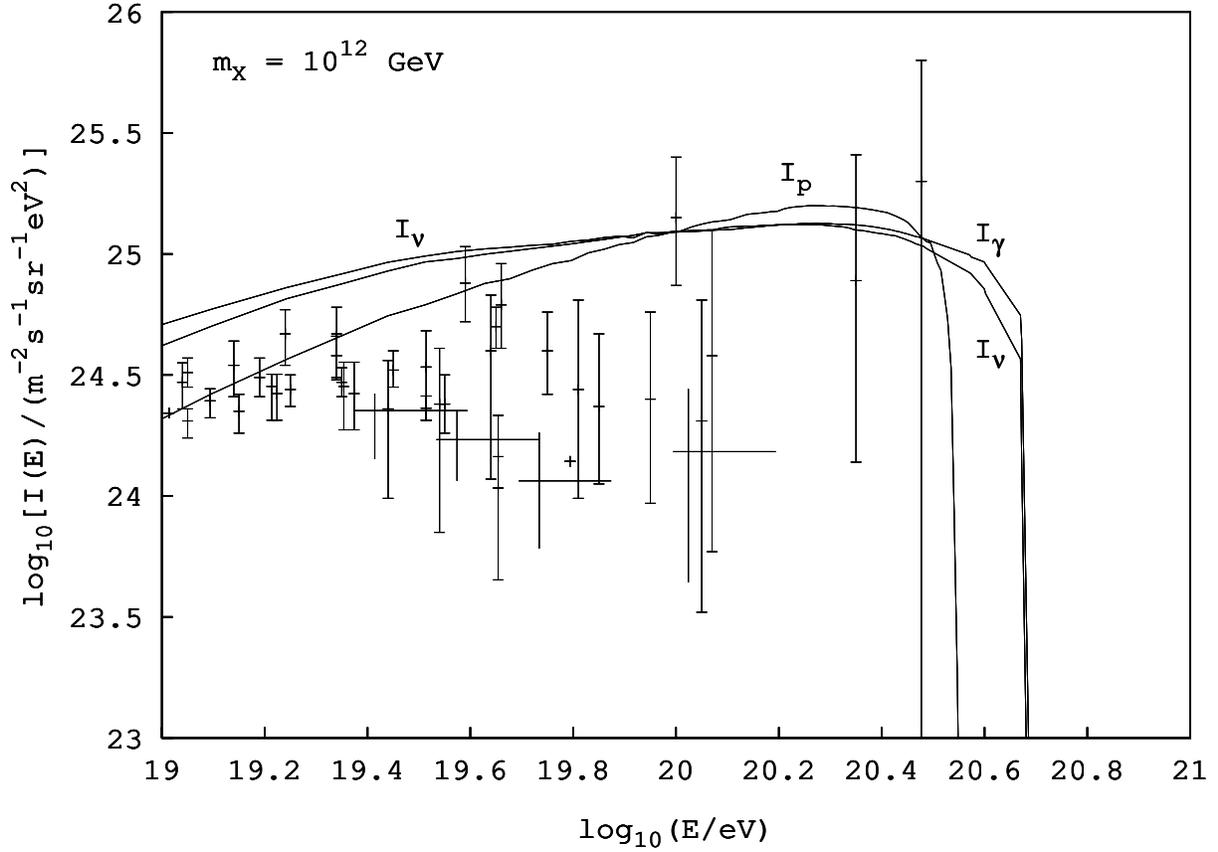}
\bigskip
\caption{Expected fluxes of protons, neutrinos and photons from decays
of a decaying particle with mass $m_X=10^{12}\gev$ (normalized as in
Figure~\protect\ref{fig7}) compared with the observations. Note that
the photon flux will be degraded through interactions with the CMB
during travel to Earth and is shown for illustrative purposes only.}
\label{fig8}
\end{figure}

\begin{figure}[t]
\epsfxsize\hsize\epsffile{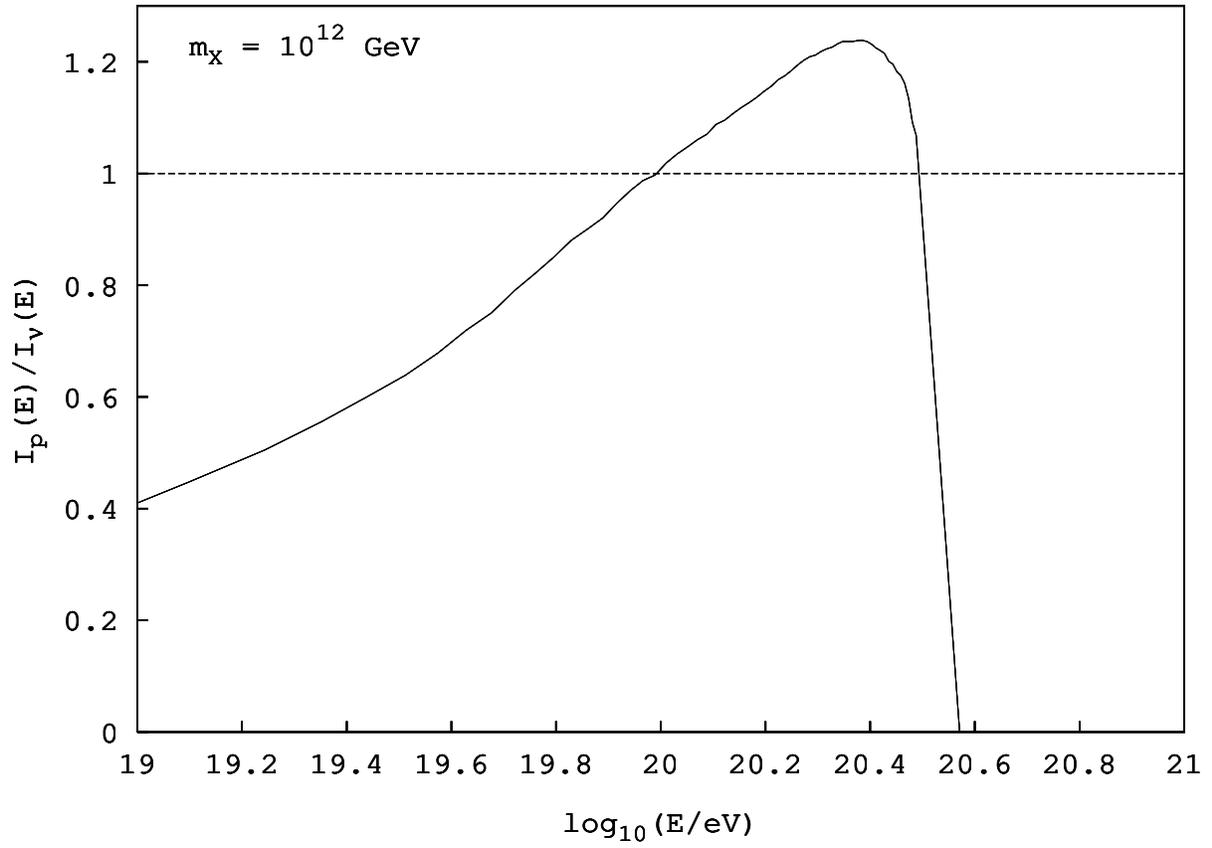}
\bigskip
\caption{The ratio of the proton to neutrino flux for the decaying
particle mass $m_X=10^{12}\gev$.}
\label{fig9}
\end{figure}

An interesting signature for forthcoming experiments is the predicted
ratio of the proton to neutrino flux \cite{watson}. In
Figure~\ref{fig8} we compare the expected flux of protons and
neutrinos for $m_X=10^{12}\gev$. (We also show the photon flux to
illustrate the difference from the prediction in ref.\cite{bkv97} but
emphasize that this will be degraded through interactions with photon
backgrounds during travel to Earth.) As can be seen, the neutrino flux
exceeds the proton flux for $10^{19}\ev\lesssim\,E\lesssim10^{20}\ev$
and also for $E\gtrsim3\times10^{20}\ev$, as may have been anticipated
from the comparison of their respective fragmentation functions. Thus
the ratio $I_p/I_\nu$ has a characteristic peak at about
$2\times10^{20}\ev$ as shown in Figure~\ref{fig9}. This could be an
useful diagonistic of the decaying particle hypothesis for future
experiments such as the Pierre Auger Project. Note that taking the
extragalactic contribution into account would boost the neutrino flux
by a factor of $\sim2$ over that shown in the figures.

\begin{figure}[t]
\epsfxsize\hsize\epsffile{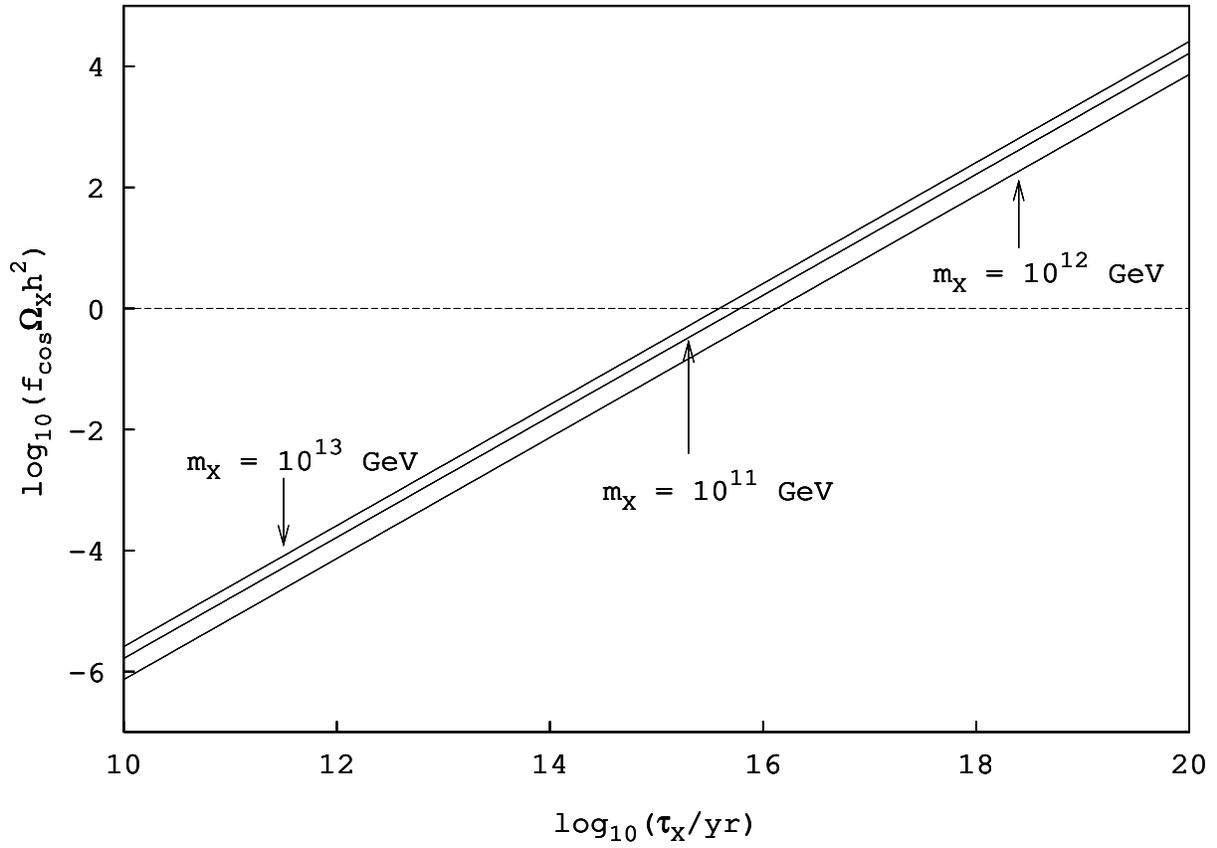}
\bigskip
\caption{The relic decaying particle abundance versus lifetime for
various masses, as required by the flux normalization to the
observations.}
\label{fig10}
\end{figure}

The abundance and lifetime of decaying particles such as cryptons are
related through the spectrum normalization (\ref{FEnorm}) as:
\begin{equation}
 \log_{10}(f_\cos \Omega_X h^2) 
 = k + \log_{10}\left(\frac{\tau_X}{{\yr}}\right)
\end{equation}
where $k = -15.78, -16.13, -15.59$ for crypton masses $m_X=10^{11},
10^{12}, 10^{13}\gev$ respectively. For a given crypton mass, a higher
lifetime must be compensated for by a higher relic abundance, as
illustrated in Figure~\ref{fig10}. So for example, if
$f_\cos\Omega_X\,h^2\sim1$, cryptons with a mass of $m_X=10^{12}\gev$
are required to have a lifetime of $\tau_X\sim10^{16}\yr$ if they are
to explain the EHECR flux. If the enhancement in the halo is
$f_\cos\sim3\times10^4$ as expected for cold dark matter, then the
lifetime may be increased to $\sim4\times10^{20}\yr$ if
$\Omega_X\,h^2\sim1$; alternatively, for the same lifetime one could
tolerate a lower relic abundance $\Omega_X\,h^2\sim3\times10^{-5}$.

With regard to the fluxes of electrons and photons, both species would
generate electromagnetic cascades on the prevalent radiation
backgrounds through pair production and inverse Compton-scattering. A
thorough analysis of such propagation effects and the resulting
modifications of the injected photon and electron spectra has been
performed \cite{ps96}. It was found that the relic decaying particles
with $m_X\gtrsim10^{14}\gev$ would contribute excessively to the
diffuse $\gamma$-ray background and are therefore ruled out. Hence,
the mass range we favour,
viz. $10^{11}\lesssim\,m_X/\gev\lesssim10^{13}$, does not lead to any
conflict with observations. This conclusion is strengthened by the
fact that according to our calculations the previous estimate
\cite{bkv97} of the $\gamma$-ray flux from decaying particles was too
high. Although the positrons released in the decays may be accumulated
in the galactic halo, the astrophysical uncertainty in the containment
time does not allow a restrictive constraint to be derived from limits
on the positron flux in cosmic rays \cite{gondolo}.

With regard to the neutrino background, the predicted flux at high
energies is well below the upper limits derived from consideration of
horizontal air showers \cite{has}, again because the decaying crypton
mass is restricted to be less than about $10^{13}\gev$. It is also
interesting to consider the flux at lower energies of ${\cal
O}(10^3)\gev$ where experiments such the forthcoming ANTARES detector
\cite{antares} will be most sensitive. As seen in Figure~\ref{fig8}
the predicted neutrino flux dominates over the proton flux at low
energies, thus the bulk of the energy released by the decaying
cryptons ends up as neutrinos.
\footnote{This expectation motivated the study undertaken earlier in
which the abundance/lifetime of massive metastable relic particles was
constrained using experimental limits on the high energy neutrino flux
set by underground nucleon decay detectors and the Fly's Eye
experiment \cite{ggs93}.} Therefore we expect the neutrino flux at TeV
energies to be at least $\sim10^8$ times larger than the EHECR flux at
$10^{20}\ev$. Moreover the neutrinos should be well correlated in both
time and arrival direction with the cosmic rays since the path length
in the galactic halo is $\lesssim100\kpc$. This is in contrast to the
case of other suggested cosmologically distant sources such as
gamma-ray bursts where the relative time delay can be upto
$\sim10^3\yr$ \cite{grbnu}.

\section{Conclusions}\label{concl}

We have investigated the hypothesis that the highest energy cosmic
rays, in particular those observed beyond the GZK cutoff, arise from
the decay of massive metastable relic particles which constitute a
fraction of the dark matter in the galactic halo. To simplify
computations (using the HERWIG Monte Carlo event generator) we have
considered only decays into $q\bar{q}$ pairs with unit branching
ratio. Comparison with experimental data indicates that a decaying
particle mass of ${\cal O}(10^{12})\gev$ is required to fit the
spectral shape while the absolute flux requires a lifetime of ${\cal
O}(10^{20})\yr$ if such particles contribute the critical density. The
predicted decay spectra may be somewhat altered if 3-body decays and
other final states (e.g. supersymmetric particles \cite{bk97}) are
considered. However our conclusions regarding the preferred mass and
relic abundance/lifetime of the decaying particle are unlikely to be
affected. In particular it would appear that the approximations used
to calculate the particle spectra in previous studies of decaying
topological defects \cite{td} and hypothetical massive particles
\cite{bkv97} were not sufficiently accurate. Our work indicates that
the topological defect model is disfavoured unless the mass of the
decaying gauge bosons is less than about $10^{13}\gev$, which is well
below the unification scale of $\sim10^{16}\gev$. (A similar
conclusion is arrived at by independent arguments in
refs.\cite{bss97,vah98}.) By contrast, cryptons from the hidden
sector of supersymmetry breaking have a mass of the required order, as
well as a decay lifetime which is naturally suppressed. However their
relic abundance is difficult to estimate reliably, although we have
argued that it may be cosmologically interesting.

The primary intention of this work is to attempt to quantify the
decaying particle hypothesis in a manner which is of interest to
experimentalists. We have therefore computed the expected neutrino to
proton ratio as a function of energy since this is an important test
of competing hypotheses for forthcoming experiments, in particular the
Pierre Auger project \cite{auger}. Of course our cleanest prediction
is that the cosmic ray spectrum should cut off just below the mass of
the decaying crypton, at $\sim3\times10^{20}\ev$. Moreover, with
sufficient event statistics it should be possible to identify the
small anisotropy which should result from the distribution of the
decaying particles in the Galactic halo \cite{dt98}. Thus although the
hypothesis investigated here is very speculative, it is nevertheless
testable. Perhaps Nature has indeed been kind to us and provided a
spectacular cosmic signature of physics well beyond the Standard
Model.

\acknowledgments{We are grateful to Mike Seymour for his help with
running HERWIG and Brian Webber for advice. We thank Martin Moorhead
for emphasizing the implications for neutrino detection and Venya
Berezinsky, M. Nagano and Alan Watson for helpful
correspondence. M.B. acknowledges financial support from the
Fellowship HSP III of the German Academic Exchange Service (DAAD).}


\end{document}